# PIONIC RADIOACTIVITY AS NEW MODE OF NUCLEAR FISSION

D. B. Ion[1,3], M. L. D. Ion[2], Reveica Ion-Mihai[2]

[1] *Institute for Physics and Nuclear Engineering, IFIN-HH, Bucharest Romania*

[2] *Bucharest University, Faculty of Physics, Bucharest, Romania*

[3] *Academy of Romanian Scientists (AOSR)*

## Abstract

In this paper a short review of the theoretical problems of the pionic radioactivity as a new nuclear mode is presented. The essential theoretical and experimental results obtained in the 25 years from the prediction of the nuclear pionic radioactivity are reviewed. Using the fission-like model it was shown that most of the SHE-nuclei lie in the region where the pionic fissility parameters attain their limiting value $X_{\pi F} = 1$ (see Fig.2-3). Hence, the SHE-region is characterized by the absence of a classical barrier toward spontaneous pion emission. Consequently, both decay modes, the pionic fission and the spontaneous fission of SHE nuclides, essentially will be determined only by shell effects. Then, it was seen that the usual predicted SHE-island of stability around the double magic nucleus 298-[114], which is not confirmed experimentally, can be explained by the dominant pionic radioactivity of the SHE-nuclei from this region. The bimodal symmetric (see Figs. 5ab) as well as, the supergiant radioactive halos (see Fig. 6) as two important signature of the nuclear pionic fission are evidentiated.

## 1. Introduction

The traditional picture of the nucleus as a collection of neutrons and protons bound together via the strong force has proven remarkable successful in understanding a rich variety of nuclear properties. However, the achievement of modern nuclear physics that not only nucleons are relevant in the study of nuclear dynamics but that pions and the baryonic resonances like $\Delta$'s and N* play an important role too. So, when the nucleus is explored at short distance scales the presence of short lived subatomic particles, such as the pion and delta, are revealed as nuclear constituents. The role of pions, deltas, quarks and gluons in the structure of nuclei is one of challenging frontier of modern nuclear physics. This modern picture of the nucleus bring us to the idea [1] to search for new exotic natural radioactivities such as: ($\pi, \mu, \Delta, N^*$)-emission during the nuclear fission in the region of heavy and superheavy nuclei. So, in 1985, we initiated the investigation of the nuclear spontaneous pion emission as a new possible nuclear radioactivity called *nuclear pionic radioactivity* (NPIR) [see Refs [1-48]] with possible essential contributions to the instability of heavy and superheavy nuclei [see [16-19, 29]]. Moreover, new exotic radioactivities such as new mode of nuclear fission (*deltonic fission* and *hyperfission*), was also suggested and investigated.

In this paper we present a short review not only of the main theoretical problems of the NPIR but also of the essential experimental results obtained in these 25 years of existence of the nuclear pionic radioactivity. Moreover, in this paper by using the recent results on the spontaneous fission half lives $T_{SF}$ of the heavy nuclei with $Z \geq 100$ we present new prediction on the pionic yields in the region of superheavy elements.

## 2. Description of quantum decay on the base of symmetry and optimality principles

It is known that many localized nuclear or subnuclear particle of finite mass should be "unstable", since the decay into several smaller particles provides many more ways to distribute the energy, and thus would have higher entropy. In quantum physics of the decay processes many physicists have adopted the description based on the <u>*totalitarian principle*</u> for this situation. It might be stated as: "*everything which is not forbidden, is compulsory*". Physicist Gell-Mann used this expression to describe the state of particle physics in the mean time he was creating the Eightfold Way, a precursor to the quark-model of hadrons. The statement is in reference to a surprising feature of particle interactions: <u>*that any interaction which is not forbidden by a small number of simple conservation laws is not only allowed*</u>, but *must* be included in the sum over all "paths" which contribute to the outcome of the interaction. Hence if it isn't forbidden, there is some probability amplitude for it to happen.

From this point of view, any decay process of the nuclear or subnuclear particles which is expected but not observed must be prevented from occurring by some conservation laws. In the spirit of this totalitarian principle we introduced [1] and started the investigation of the *spontaneous pion emission during the fission* as a new nuclear radioactivity called *pionic radioactivity*.

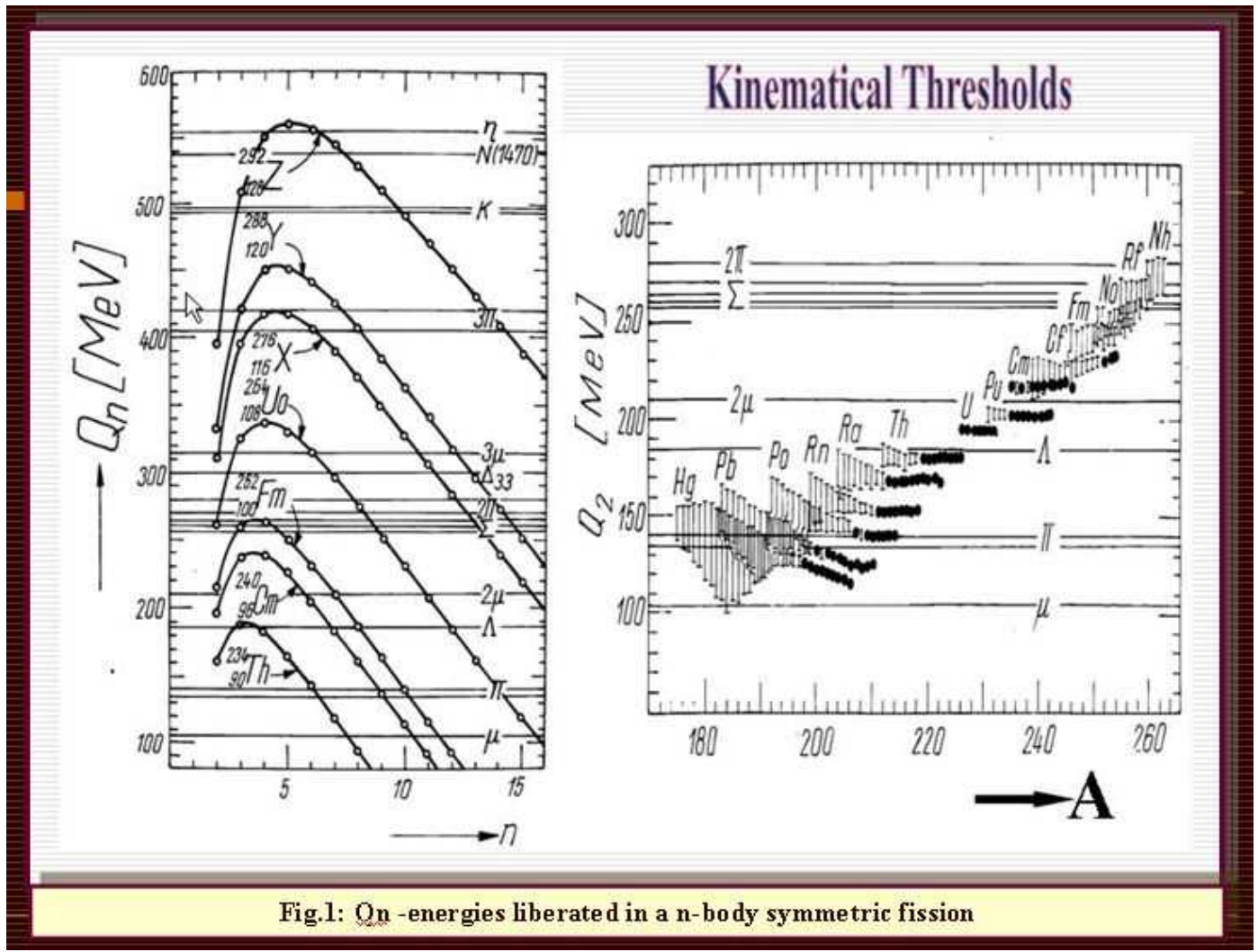

Fig.1: $Q_n$ -energies liberated in a n-body symmetric fission

### 3. Investigation of the lepton and meson emission during spontaneous fission

We started our investigations the values of energies liberated in n-body fission of heavy nuclei. So, in Fig.1 we presented the values of energies liberated in n-body symmetric fission as well as in two-body fission of heavy nuclei. Then, the kinematical thresholds for different particle emission during fission are indicated by orizontal lines.

These investigations was conducted in agreement with a special "fission model", for each x-particle emitted during fission, similar with that presented in our paper [20] for the pion emission during fission. A fission-like model [5, 20, 29] for the pionic radioactivity was regarded as a first stage in the development of an approximate theory of this new phenomenon in order to takes into account the essential degree of freedom of the system: $\pi$ – *fissility*, $\pi$ – *fission barrier height*, etc.

Therefore, let us consider

$$E_C^{\pi F}(0) = E_C^0 - \alpha m_\pi \tag{1}$$

$$E_S^{\pi F}(0) = E_S^0 - (1-\alpha)m_\pi \tag{2}$$

where $E_C^0$ and $E_S^0$ are the usual Coulomb energy and surface energy of the parent nucleus given by

$$E_C^0 = \gamma Z^2 / A^{1/3} \quad \text{and} \quad E_S^0 = \beta A^{2/3} \tag{3}$$

with $\beta = 17.80$ MeV and $\gamma = 0.71$ MeV. $\alpha$ is a parameter defined so that $\alpha m_\pi$ and $(1-\alpha)m_\pi$ are the Colombian and nuclear contributions to the pion mass (for $\alpha = 1$, the entire pion mass is obtained only from Coulomb energy of the parent nucleus). So, by analogy with binary fission was introduced the pionic fissility $X_{\pi F}$ which is given by

$$X_{piF}^{(\alpha)} = \frac{E_C^{\pi F}(0)}{2 E_S^{\pi F}(0)} = \frac{1}{2} \cdot \frac{E_C^0 - \alpha m_\pi}{E_S^0 - (1-\alpha)m_\pi}, \quad 0 \leq \alpha \leq 1 \tag{4}$$

or

$$\left(\frac{Z^2}{A}\right)_{\pi F} = \frac{Z^2}{A} - \frac{m_\pi}{\gamma A^{2/3}} \cdot \frac{\alpha - (1-\alpha)E_C^0/E_S^0}{1-(1-\alpha)m_\pi/E_S^0} \tag{5}$$

In the particular case $\alpha = 1$ we have

$$X_{\pi F} = X_{SF} - \frac{m_\pi}{2 E_S^0} \quad \text{or} \quad \left(\frac{Z^2}{A}\right)_{\pi F} = \frac{Z^2}{A} - \frac{m_\pi}{\gamma A^{2/3}} \tag{6}$$

Therefore, in these xF-fission models we introduced the essential parameters of these new types of radioactivitis such as: $Q_{xF}$-*values*, *xF-fissility parameters*, *xF-fission barrier* (or kinematical thresholds) (see Figs 1-3), nuclear configuration of the parent nucleus at xF-saddle points, etc. Some of these results are presented in the above last picture especially for the pions and muons during spontaneous fission (see Figs.4-5). Moreover, on the basis of the principle of detailed equilibrium [5], the probabilities of xF-fissions as functions of temperature at "saddle points" was also obtained.

We note that the above definitions are valid only with the constraints

$$E_C^{\pi F}(0) + E_S^{\pi F}(0) + m_\pi = E_C^0 + E_S^0 \tag{7}$$

They are applicable to all kind of exotic nuclear and subnuclear decays with the substitution:

$$m_\pi \Rightarrow \Delta m_x \tag{8}$$

where $\Delta m_x$ is the energy necessary to create the x-particle on mass shell.
The *dynamical thresholds* for the pionic fission are obtained just as in fission case by using the substitution:

$$X_{SF} \to X_{\pi F} \tag{9}$$

In Fig. 2 we presented the regions from the plane (A, Z) in which some parent nuclei are able to emit spontaneously pions (or muons and neutrino) during the nuclear spontaneous fission.

Therefore, according to Fig. 2a, we have the following important regions:

- SHE (super heavy elements)-region, indicated by white circles, where $X_{\pi F} > 1$ and $Q_\pi > 0$, all the nuclei are able to emit spontaneously pions during the SF since no pion fission barrier exists.
- HE (heavy elements)-zone marked by signs plus (+++), corresponding to $X_{\pi F} < 1$ and $Q_\pi > 0$, where all the nuclei can emit spontaneously pion only by quantum tunneling of the pionic fission barrier.
- E-region, indicated by signs minus (----) where the spontaneous pion emission is energetically interdicted since $Q_\pi < 0$.

# Physical Regions for Pionic Radioactivity

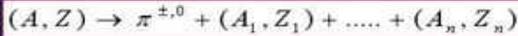
$(A,Z) \to \pi^{\pm,0} + (A_1, Z_1) + \ldots + (A_n, Z_n)$

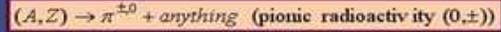
$(A,Z) \to \pi^{\pm,0} + anything$ (pionic radioactivity $(0,\pm)$)

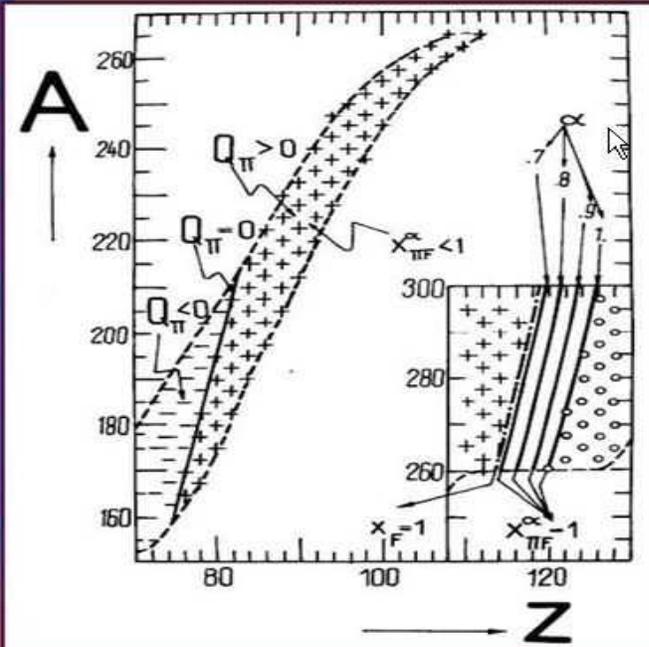

In a pionic fission-like model we defined $\pi$-fissility parameter

$$X_{\pi F}^{\alpha} = \frac{E_C^{\alpha}(0)}{2 E_S^{\alpha}(0)} = \frac{E_C^0 - \alpha m_{\pi}}{2[E_S^0 - (1-\alpha) m_{\pi}]},$$

or

$$\left(\frac{Z^2}{A}\right)_{\pi F} = \frac{Z^2}{A} - \frac{m_{\pi}}{\gamma A^{1/3}} \frac{\alpha - (1-\alpha) E_C^0 / E_S^0}{1 - (1-\alpha) m_{\pi} / E_S^0}$$

where

$E_C^{\alpha}(0) = E_C^0 - \alpha m_{\pi}$, $E_S^{\alpha}(0) = E_S^0 - (1-\alpha) m_{\pi}$,
$E_C^0 = \gamma Z^2 / A^{1/3}$, $E_S^0 = \beta A^{2/3}$,
$\gamma = 0.71$ MeV, $\beta = 17.80$ MeV, $0 \le \alpha \le 1$

The regions from plane (A,Z) where the parent nuclei are able to emit pions by tuneling a pionic-fission barrier (+++) and without the pi-fission barrier (ooo).

Fig. 2: Physical regions in the plane (A,Z) for pionic fission

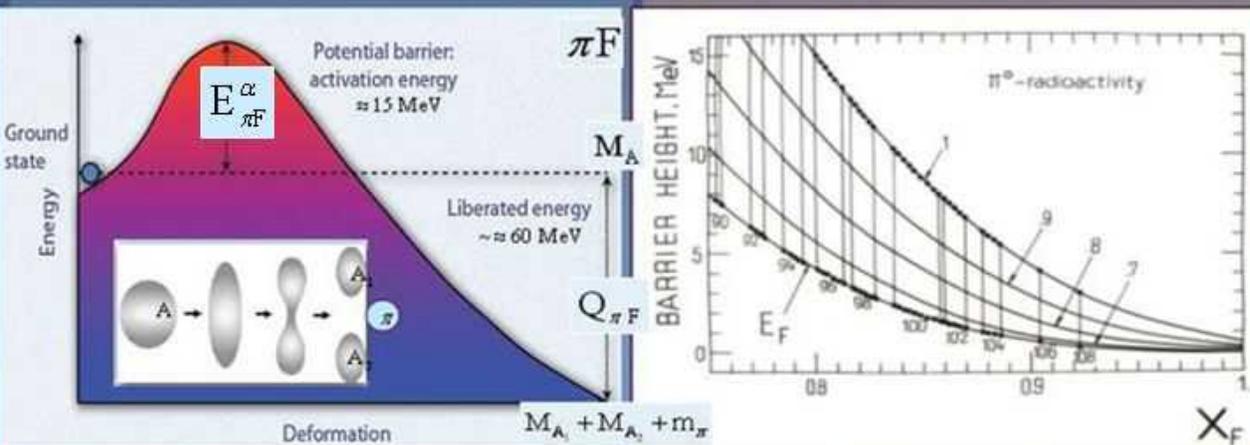

$$E_{\pi F}^{\alpha}(LD) = (E_{\pi F}^{\alpha})_S^0 [\, 0.73(1 - X_{\pi F}^{\alpha})^3 - 0.33(1 - X_{\pi F}^{\alpha})^4 + 1.92(1 - X_{\pi F}^{\alpha})^5 - 0.21(1 - X_{\pi F}^{\alpha})^6 ]$$

where:

$$(E_{\pi F}^{\alpha})_S^0 = E_s^0 - (1-\alpha) m_{\pi}, \quad (E_{\pi F}^{\alpha})_C^0 = E_C^0 - \alpha m_{\pi}$$

and

$$X_{\pi F}^{\alpha} = [E_C^0 - \alpha m_{\pi}] / 2[E_S^0 - (1-\alpha) m_{\pi}]$$

Fig. 3: Schematic description of pionic fission barrier (left) and liquid drop (LD)-model predictions for the barrier height of neutral pion emission during two-body fission

Next, by analogy with binary fission the *barrier height* for the pionic fission in a liquid drop model can be written as in Fig.3 while the nuclear configuration at the saddle point is given by

$$R^{\alpha}(\theta) = \frac{R_0}{\lambda}[1 + \varepsilon_2^{\alpha} \cdot P_2 \cos(\theta) + \varepsilon_4^{\alpha} \cdot P_4 \cos(\theta) + \varepsilon_6^{\alpha} \cdot P_6 \cos(\theta) + ...] \qquad (10)$$

where

$$\varepsilon_2^{\alpha} = 2.33(1 - X_{\pi F}^{\alpha}) - 1.23(1 - X_{\pi F}^{\alpha})^2 + 9.5(1 - X_{\pi F}^{\alpha})^3 - 8.05(1 - X_{\pi F}^{\alpha})^4 + ... \qquad (11a)$$

$$\varepsilon_4^{\alpha} = 1.98(1 - X_{\pi F}^{\alpha})^2 - 1.70(1 - X_{\pi F}^{\alpha})^3 + 17.74(1 - X_{\pi F}^{\alpha})^4 + ... \qquad (11b)$$

$$\varepsilon_6^{\alpha} = 0.95(1 - X_{\pi F}^{\alpha}) + ... \qquad (11c)$$

$R_0$ is the spherical radius and $\lambda$ is a scale factor just as in binary spontaneous fission.
Next, by using Eqs. (10)-(11), we obtained the predictions presented in Fig. 4 for the pionic yields as well as for the configurations at the sadle points of the pionic radioactivity.

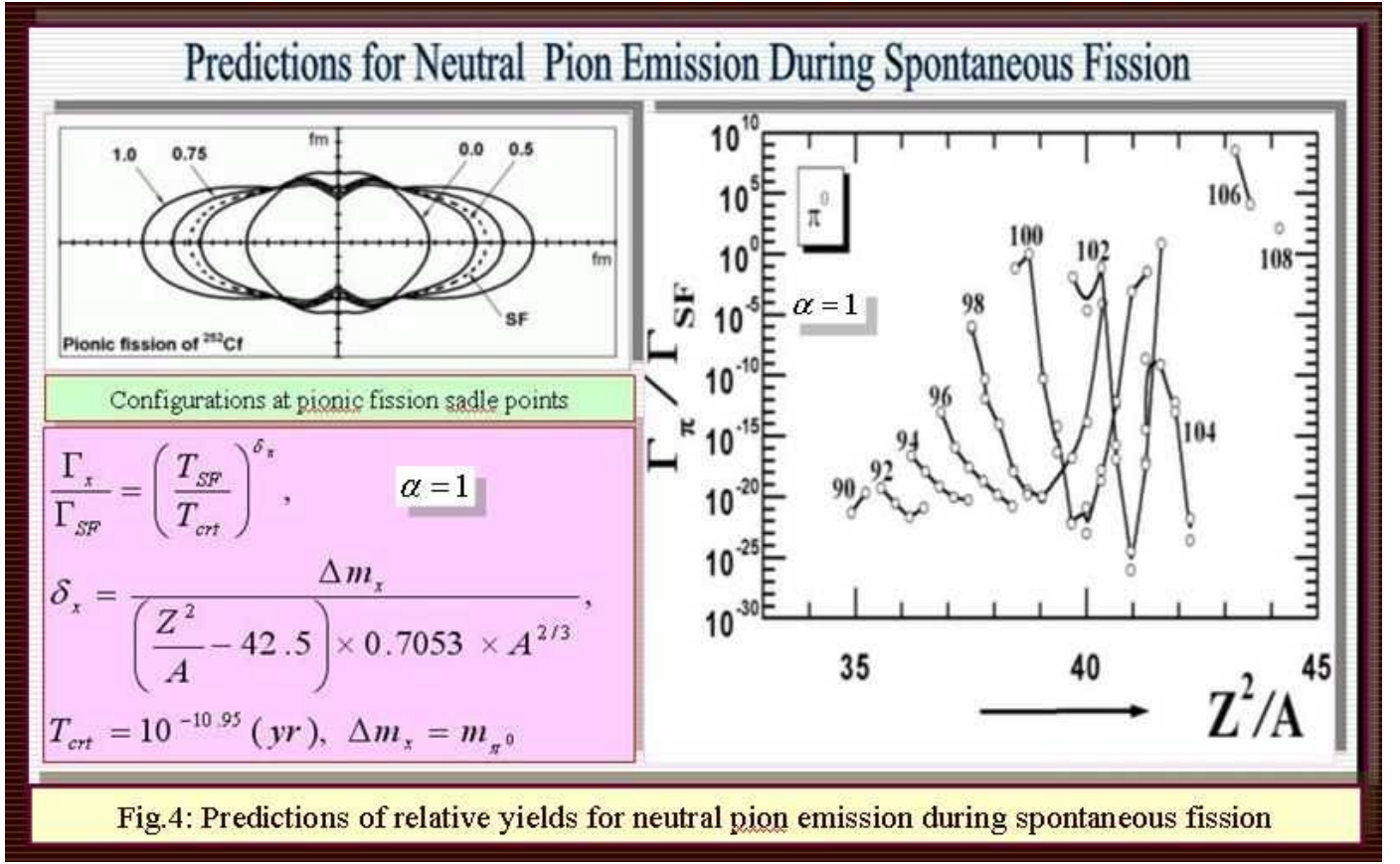

Fig.4: Predictions of relative yields for neutral pion emission during spontaneous fission

Detailed predictions for the relative yields of the pionic radioactivity are obtained [4-7, 20, 33] by assuming a scaling law for any kind of fission accompanied by the light particle emission.

## 4. Bimodal fission as an indirect experimental evidence for pionic radioactivity

In the paper [31,34] we presented new predictions for the spontaneous pion emission accompanied by fission for all nuclei with $100 < Z < 108$. Then, using the S-matrix unitarity diagrams (see Figs. 5a,b), we proved that the bimodal fission can be interpreted as an imporant experimental evidence for the pionic radioactivity dominance at those parent nuclei with symmetric mass distribution of the fission fragments. So, in Fig. 5a we displayed the *mass distributions of nuclear fragments for superheavy nuclei* while in Fig. 5b we presented a schematic description of spontaneous fission (SF) as well as of NPIR of 258-Fm in terms of two types of unitarity diagrams, via two step nuclear reactions, as follows

**SF – Asymmetric fission (*A*)**
first step: $^{258}Fm \rightarrow {}^{50}Ar + {}^{208}Pb$, second step: $^{50}Ar + {}^{208}Pb \rightarrow {}^{126}Sn + {}^{132}Sn$
**SF – Symmetric fission (*S*)**
first step: $^{258}Fm \rightarrow \pi + {}^{50}Ar + {}^{208}Pb$, second step: $\pi + {}^{50}Ar + {}^{208}Pb \rightarrow {}^{126}Sn + {}^{132}Sn$
**NPIR – Asymmetric fission (*A*)**
first step: $^{258}Fm \rightarrow {}^{50}Ar + {}^{208}Pb$, second step: $^{50}Ar + {}^{208}Pb \rightarrow \pi + {}^{126}Sn + {}^{132}Sn$
**NPIR – Symmetric fission (*S*)**
first step: $^{258}Fm \rightarrow \pi + {}^{50}Ar + {}^{208}Pb$, second step: $\pi + {}^{50}Ar + {}^{208}Pb \rightarrow \pi + {}^{126}Sn + {}^{132}Sn$

Then, the scenario described in Figs. 5ab, which include in a more general and exact form the idea of transition to a new phase of nuclear matter, is leading us to point out the connection between the observed bimodal symmetric SF at the nuclides $^{258}F$, $^{259}F$, $^{259}Md$, $^{260}Md$, $^{258}No$, $^{262}No$, $^{258}Rf$, $^{260}Rf$, $^{262}Rf$ with a significant high NPIR-yields predicted for these nuclei (see Table 1 in [30]).

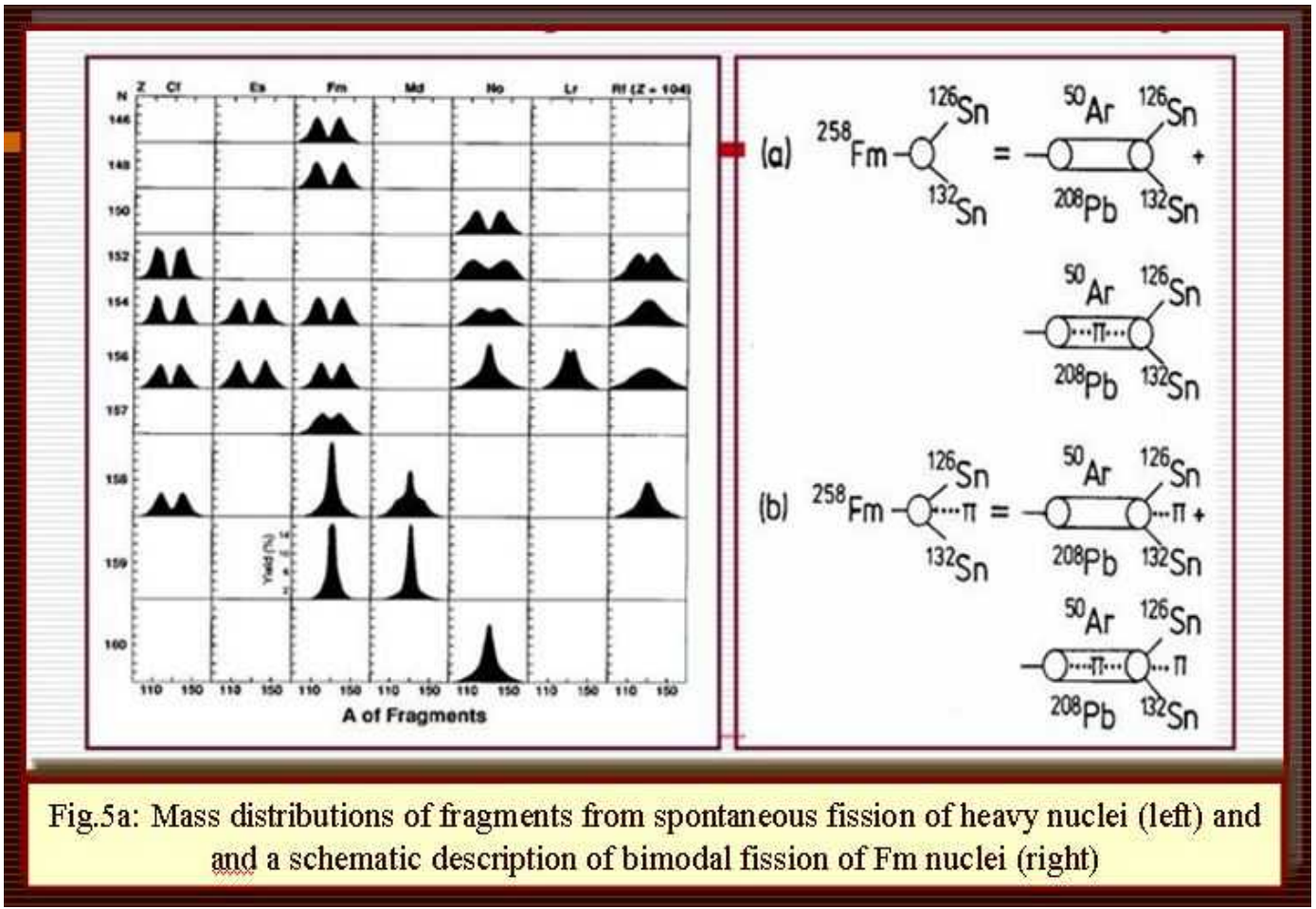

Fig.5a: Mass distributions of fragments from spontaneous fission of heavy nuclei (left) and and a schematic description of bimodal fission of Fm nuclei (right)

For sake of completeness we recall here the results [18,19,31] regarding the competition between the pionic radioactivity and the spontaneous fission of the SHE-nuclides. Indeed, using the same fission-like model, it was shown that most of the SHE-nuclei lie in the region where the pionic fissility parameters attain their limiting value $X_{\pi F}=1$. Hence, the SHE-region is characterized by the absence of a classical barrier toward spontaneous pion emission (see Figs.2-3). Consequently, both decay modes pionic fission and the spontaneous fission of SHE nuclides essentially will be determined only by shell effects. Then, it was seen that the usual predicted SHE-island of stability around the double magic nucleus 298-[114], which is not confirmed experimentally, can be explained by the dominant pionic radioactivity of the SHE-nuclei from this region. Therefore, we believe that in future searches for transfermium as well as for SHE-nuclei, the pi-

*fission* as detection method [45] can play an essential role in the discovery of this new phase of nuclear matter produced by the presence of pions on mass shell in the nuclear medium.

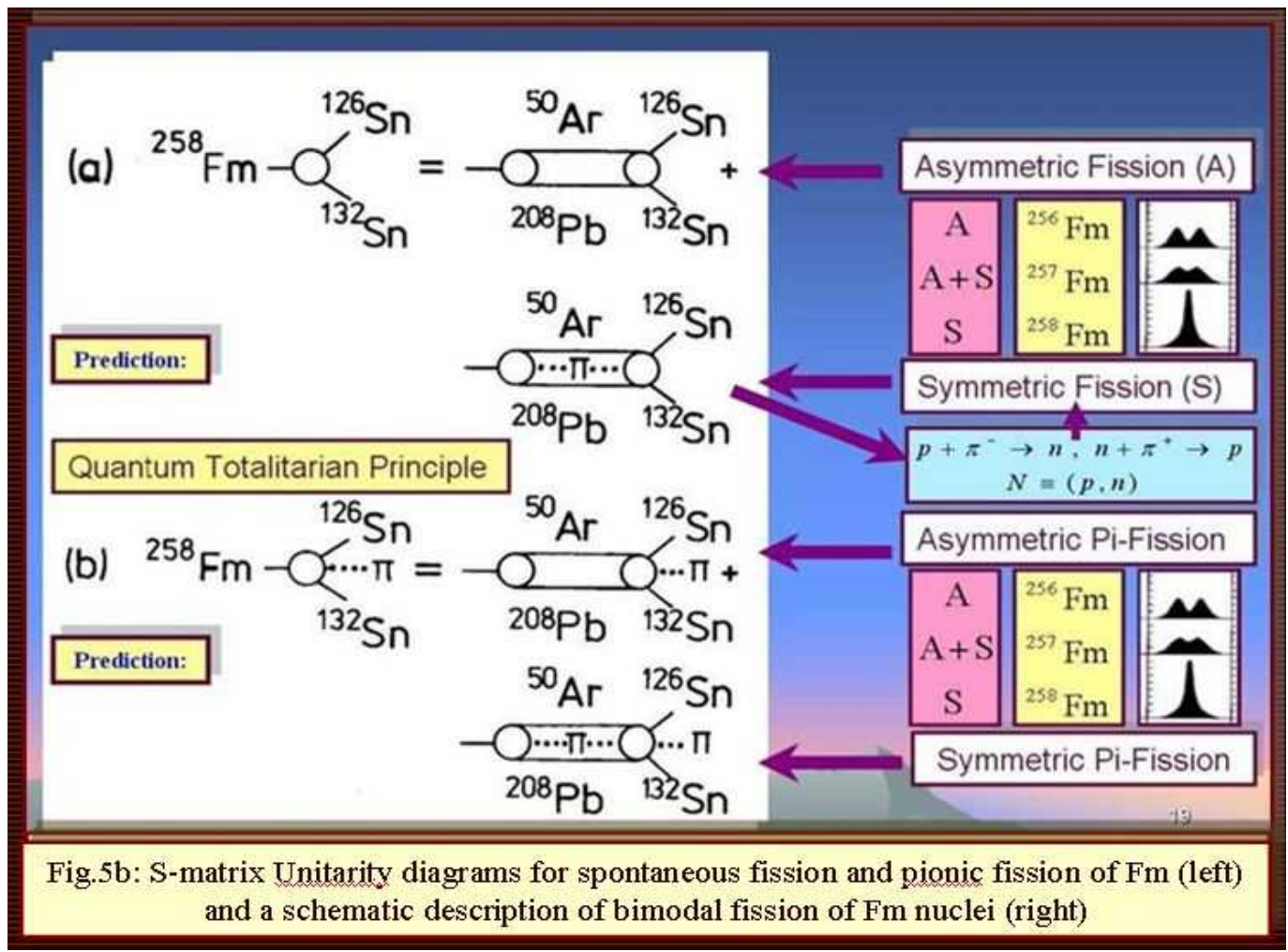

Fig.5b: S-matrix Unitarity diagrams for spontaneous fission and pionic fission of Fm (left) and a schematic description of bimodal fission of Fm nuclei (right)

## 5. Supergiant radioactive halos as integral record of pionic radioactivity

The radioactive halos were first reported between 1880 and 1890 and their origin was a mystery until the discovery of radioactivity and its power of coloration. Therefore, aside from their interest as attractive mineralogical oddities, the radioactive halos are of great interest for the nuclear physics because they are an integral record of radioactive decay in minerals. This integral record is detailed enough to allow estimation of the energy involved in the decay process and to identify the decaying nuclides through genetic connections. This latter possibility is particularly exciting because there exist certain classes of halos, such as the dwarf halos, the X-halos, the giant halos and the supergiant halos, which cannot be identified with the ring structure of the known alpha-emitters. Hence, barring the possibility of a non radioactive origin, these new variants of halos can be interpreted as evidences for hitherto undiscovered alpha-radionuclide, as well as, signals for the existence of new types of radioactivities. In the spirit of these ideas based on the pionic radioactivity hypothesis in the paper [30] the pionic radiohalos (PIRH), as the integral record in time of the pionic nuclear radioactivity of the heavy nuclides with $Z>80$ from the inclusions from ancient minerals, are proposed. Then, it is shown that the essential characteristic features of the observed supergiant halos (SGH) [see Fig. 9 (left)] observed by Grady and Walker [51] and Laemmlein [52], are reproduced with a surprising high accuracy by those of the pionic radiohalos.

# 6. Conclusions

The experimental and theoretical results obtained on the pionic radioactivity in more than 25 years can be summarized as follows:

1. A fission-like model (see Ref.[20]) for the pionic radioactivity was regarded as a first stage in the development of an approximate theory of this new phenomenon that takes into account the essential degree of freedom of the system: *$\pi$–fissility, $\pi$–fission barrier height, nuclear configuration at $\pi F$ – saddle point,* etc. Detailed predictions for the pionic radioactivity are obtained in the papers (see Ref. [4-7,29,33]) by assuming a scaling law of all kind of fissions accompanied by the light particle emission.
2. A statistical model for NPIR as well as a detailed balance model for pionic radioactivity are presented in our papers [8,9].
3. A new interpretation of the experimental bimodal fission of the heavy and superheavy nuclides, in terms of the S-matrix unitarity diagrams, is obtained in the papers. [31,34] So, the presence of the symmetric mode in the fragment mass-distribution at transfermium nuclei can be interpreted as experimental signature of the pionic radioactivity.
4. Since $\pi$–*fissility* $X_{\pi F}=1$ in the region of superheavy elements(SHE) (see Fig.2) it is expected that this new degree of nuclear instability as is the pionic radioactivity to becomes dominant [16-19]. Then, the predicted SHE-island of stability around the double magic nucleus 298-[114], which is not confirmed experimentally, can be explained by the dominant pionic radioactivity in this region.
5. The nuclear pionic radioactivity (NPIR) was experimentally investigated by many authors [23]-[28] [36]-[47]. A short review of the experimental limits obtained on the spontaneous NPIR yields is presented in Table 1. The best experimental limit for $\pi^0$ -yields has been reported for $^{252}$Cf by Bellini et al. [40]. They reached an upper limit of $3.10^{-13}$, a value close to the theoretical prediction [5].

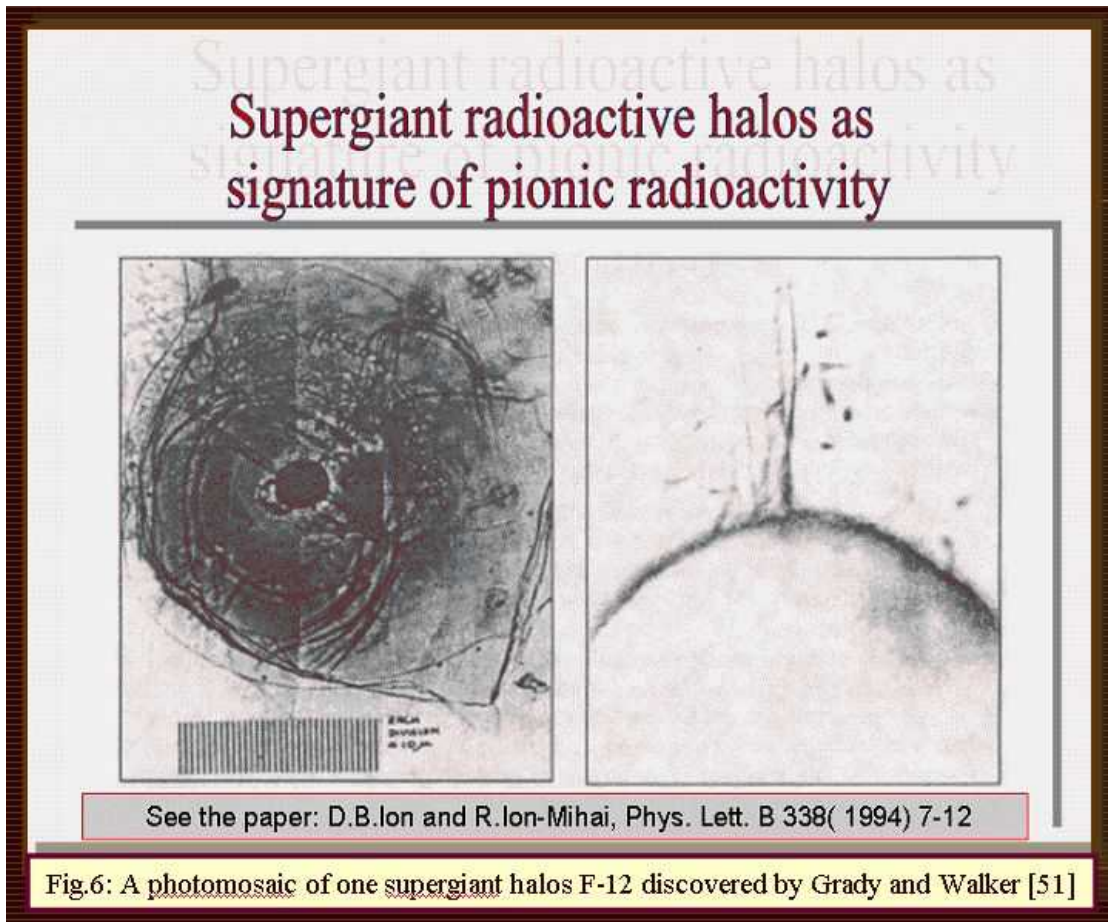

See the paper: D.B.Ion and R.Ion-Mihai, Phys. Lett. B 338( 1994) 7-12

Fig.6: A photomosaic of one supergiant halos F-12 discovered by Grady and Walker [51]

6. The unusual background, observed by Wild et al.[50] in $(E, \Delta E)$-energy region below that characteristic for long range alpha emission from $^{257}$Fm was interpreted by Ion, Bucurescu and Ion-Mihai [28] as being produced by negative pions emitted spontaneously by $^{257}$Fm. Then, they inferred value of the pionic yield is: $\Gamma_\pi/\Gamma_{SF} < (1.2 \pm 0.2) \cdot 10^{-3}$ $\pi^-$/fission. In a similar way, Janko and Povinec [43], obtained the yield $\Gamma_\pi/\Gamma_{SF} < (7 \pm 6) \cdot 10^{-5}$ : $\pi^+$/fission.
7. The pionic $\pi^\pm-$ supergiant radiohalos are introduced by us in the paper [30] (see also [32,35]). Then, it is shown [30] that the supergiant radiohalos (SGH), discovered by Grady and Walker [51] and Laemmlein [52] can be interpreted as being the $\pi^- - radiohalos$ and $\pi^+ - radiohalos$, respectively. Hence, these supergiant radiohalos can be considered as experimental evidences of the integral record in time of the natural pionic radioactivity from radioactive inclusions in ancient minerals.

Finally, we note that dedicated experiments, using nuclei with theoretically predicted high pionic yields (e.g. $^{258}$Fm, $^{259}$Fm, $^{258}$No, $^{260}$No, $^{254}$Rf, and $^{264}$Hs ), are desired since the discovery of the nuclear pionic radioactivity is of fundamental importance in nuclear science not only for the clarification of high instability of SHE- nuclei.

Table 1: Experimental results on Pionic Radioactivity

| Year | Authors | Laboratory | Parent nuclei | $\Gamma_{\pi^0}/\Gamma_{SF}$ | $\Gamma_{\pi^\pm}/\Gamma_{SF}$ |
|---|---|---|---|---|---|
| 1986 | D.B.Ion et al. Rev.Roum.Phys.31,551 | IFIN-Bucharest Romania | $^{259}$Md | | $< 10^{-5}$ |
| 1987 | D.Bucurescu et al. Rev.Roum.Phys.32, 849 | IFIN- Romania Bucharest | $^{252}$Cf | | $< 10^{-8}$ |
| 1988 | M.Ivascu et al. Rev.Roum.Phys.32, 937 | IFIN-Bucharest Romania | $^{252}$Cf | | $< 10^{-8}$ |
| 1988 | C.Cerruti et al. Z.Phys. A329, 383 | CEN-Saclay France | $^{252}$Cf | $< 10^{-8}$ | |
| 1988 | J.R.Beene et al. Phys.Rev. C38, 569 | ORNL USA | $^{252}$Cf | $< 1.5 \cdot 10^{-9}$ | |
| 1989 | J. Julien et al. Z.Phys, A332, 473 | CEN-Saclay France | $^{252}$Cf | $< 10^{-12}$ | |
| 1989 | Yu.Adamchuk et al. Sov. J. Nucl. Phys. 49,932 | I.V.Kurhatov Russia | $^{252}$Cf | | $< 5 \cdot 10^{-8}$ |
| 1989 | D.B.Ion et al. Rev.Roum.Phys. 34, 261 | IFIN-Bucharest Romania | $^{257}$Fm | | $(1.2 \pm 0.2) \cdot 10^{-3} (\pi^-)$ |
| 1989 | S.Stanislaus et al. Phys.Rev. C39, 295 | University British Columbia Canada | $^{252}$Cf | $< 3.3 \cdot 10^{-10}$ | |
| 1989 | S.Stanislaus et al. Phys.Rev. C39, 295 | University British Columbia Canada | $^{238}$U | $< 3.1 \cdot 10^{-4}$ | |
| 1991 | J.N.Knudson et al. Phys.Rev. C44, 2869 | LANL USA | $^{252}$Cf | $< 1.37 \cdot 10^{-11}$ | |
| 1991 | V.Bellini et al. [19] Proc. Bratislava Conference | CEN Saclay France | $^{252}$Cf | $< 3 \cdot 10^{-13}$ | |
| 1991 | K.Janko et al. [22] Proc. Bratislava Conference | Comenius University Czechoslovakia | $^{257}$Fm | | $< 7 \cdot 10^{-5} (\pi^+)$ |
| 1992 | H.Otsu et al. Z.Phys. A 342, 483 | Tokio University Japan | $^{252}$Cf | | $< 1.3 \cdot 10^{-8} (\pi^-)$ |
| 2002 | Khryachkov et al. Instr. Exp. Tech. 45, 615 | LIPPE Obninsk Russia | $^{252}$Cf | | Not estimated |

This research was supported by CNCSIS under contract ID-52-283/2007.

# 7. References

[1] D.B.Ion, M. Ivascu, and R. Ion-Mihai, *Spontaneous pion emission as new natural radioactivity*, Ann. Phys. (N.Y.) **171,** 237 (1986).
[2] D. B. Ion, R. Ion-Mihai and M. Ivascu, *Spontaneous pion emission as a new nuclear radioactivity*, Rev. Roum. Phys. **31,** 205 (1986).
[3] D. B. Ion, R. Ion-Mihai and M. Ivascu, *Optical theorem for inclusive decay processes and pionic nuclear radioactivity*, Rev. Roum. Phys. **32,** 1037 (1987).
[4] D. B. Ion, R. Ion-Mihai and M. Ivascu, *Theoretical predictions for nuclear pionic radioactivity*, Rev. Roum. Phys. **33,** 239 (1988).
[5] D. B. Ion, R. Ion-Mihai and M. Ivascu, *Predictions for pionic radioactivity of even-even parent nuclei*, Rev. Roum. Phys. **33,** 1071 (1988).
[6] D. B. Ion, R. Ion-Mihai and M. Ivascu, *Predictions of pionic radioactivity of heavy nuclei with A-odd*, Rev. Roum. Phys. **33**, 1075 (1988).
[7] D. B. Ion, R. Ion-Mihai and M. Ivascu, *Predictions for pionic nuclear radioactivity from fission data of even-even nuclei*, Rev. Roum. Phys. **34,** 359 (1989).
[8] D. B. Ion and R. Ion-Mihai, *A statistical model for pionic radioactivity*, Stud.Cercet.Fiz.**43,** 219 (1991).
[9] D. B. Ion , *Detailed balance model for pionic radioactivity*, Stud. Cercet. Fiz. **43,** 385 (1991).
[10] D. B. Ion and R. Ion-Mihai, *Pionic radioactvity versus spontaneous fission at transfermium nuclei*, Rom. Journ. Phys. **43,** 179 (1998).
[11] D. B. Ion and R. Ion-Mihai, *Supergiant halos as experimental evidence for the pionic radioactivty*, Rom. J. Phys**. 44,** 703 (1999).
[12] D. B. Ion, R. Ion-Mihai and M. Ivascu, *Spontaneous muon emission as new natural radioactivity*, Rev. Roum. Phys. **31,** 209 (1986).
[13] D. B. Ion, R. Ion-Mihai and M. Ivascu, *Spontaneous Lambda emission as a new natural strange radioactivity*, Rev. Roum. Phys. **33,**109 (1988).
[14] D. B. Ion, R. Ion-Mihai and M. Ivascu, *Hyperfission as a new mode of fission*, Rev. Roum. Phys. **34,** 461 (1989).
[15] D. B. Ion, A. C. D. Ion and R. Ion-Mihai, *Deltonic fission as a new mode of fission*, Roum. J. Phys. **38,** 31 (1993).
[16] D. B. Ion, R. Ion-Mihai and M. Ivascu, *Superheavy elements: Do they really form an island ?* Rev. Roum. Phys. **35**, 471 (1990).
[17] D. B. Ion, R. Ion-Mihai and M. Ivascu, *Superheavy elements: (II) The island of relative stability against of pionic decay mode*, Rev. Roum. Phys . **35,** 579 (1990).
[18] D. B. Ion, R. Ion-Mihai and M. Ivascu, *Pionic radioactivity versus spontaneous fission in decay of superheavy nuclei*, Rev. Roum. Phys. **36,** 163 (1991).
[19] D. B. Ion, R. Ion-Mihai and M. Ivascu, *Pionic radioactivity as dominant decay mode of superheavy nuclei with A-odd*, Rev. Roum. Phys**. 36,** 261 (1991).
[20] D. B. Ion, *Theoretical problems of pionic radioactivity*, Rev. Roum. Phys. **37,** 347 (1992); D.B.Ion, Reveica Ion-Mihai, M. L. D. Ion, *New Nuclear and Subnuclear Exotic Decays*, Rom.Rep.Phys. **59,** 1026 (2007).
[21] D. B. Ion R. Ion-Mihai and M. Ivascu, *Experimental Evidence for spontaneous pion emission from the ground state of transuranium nuclides*, Rev. Roum. Phys**. 31,** 551(1986).
[22] D. B. Ion, R. Ion-Mihai and M. Ivascu, *Experimental evidences for spontaneous pion emission from the ground states of 252-Cf* , Rev. Roum. Phys. **32,** 299 (1987).
[23] D. Bucurescu, M. Brehui, M. Haiduc, D. B. Ion, R. Ion-Mihai, C. Petrascu, M. Petrascu, K. Tolstov and V. Topor, *Search for spontaneous pion emission in Cf-252* Rev. Roum. Phys. **32**, 849 (1987).
[24] M. Ivascu, M. Brehui, M. Haiduc, D. B. Ion, R. Ion-Mihai, C. Petrascu, M. Petrascu, K. Tolstov and V. Topor, *Search for pion emission in thermal induced fission of U-235*, Rev. Roum. Phys**. 33,** 105 (1988).

[25] M. Ivascu, D. B. Ion, D. Bucurescu, D. Cutoiu, D. Galeriu, R. Ion-Mihai and N. Zamfir, *Search for spontaneous emission of mesons in Cf-252*, Rev. Roum. Phys. **32,** 937 (1987).

[26] D. B. Ion, M. Haiduc, R. Ion-Mihai, M.Ivascu, M.Petrascu, K. Tolstov and V. Topor, *The net pionic background near VVRS reactor*, Rev. Roum. Phys. **36,** 373 (1991).

[27] V. Topor-Pop, D. B. Ion, M. Ivascu, R. Ion-Mihai, M. Petrascu, M. Haiduc and K. Tolstov, *Search for spontaneous emission of pions*, Rev. Roum. Phys. **37,** 211(1992).

[28] D. B. Ion, R. Ion-Mihai and D. Bucurescu, *Possibility of pions emitted in spontaneous fission of Fermium 257*, Rev. Roum. Phys. **34,** 261 (1989).

[29] D. B. Ion, *Theoretical problems of pionic radioactivity*, Europhysics Conf. Abstracts (Ed. K. Bethge, Frankfurt/M) 14-th Europhysics Conf. on Nuclear Physics : Rare Nuclear Decays and Fundamental Processes, Bratislava, October 22-26, 1990, p.48.

[30] D.B.Ion and R. Ion-Mihai, *Is supergiant halos experimental evidence for the pionic radioactivity?*, Phys. Lett. **B 338,** 7 (1994).

[31] D.B.Ion and R. Ion-Mihai, *Pionic radioactivity versus spontaneous fission at transfermium nuclei*, Rom. J. Phys. **43,** 179 (1998).

[32] D.B.Ion and R. Ion-Mihai, *Supergiant halos as experimental evidence for the pionic radioactivity*, Rom. J. Phys. **44,** 703 (1999).

[33] D.B.Ion, Reveica Ion-Mihai, M. L. D. Ion and Adriana Sandru, *Spontaneous pion emission during fission. A new nuclear radioactivity,* Rom. Rep. Phys. **55,** 571 (2003).

[34] D.B.Ion, Reveica Ion-Mihai, M. L. D. Ion and Adriana Sandru, *Is bimodal fission an indirect experimental evidence for pionic radioactivity?,* Rom. Rep. Phys**. 55,** 557 (2003).

[35] D.B.Ion, Reveica Ion-Mihai, M. L. D. Ion and Adriana Sandru, *Supergiant halos as integral record of natural nuclear pionic radioactivity,* Rom. Rep. Phys. **55,** 565 (2003).

[36] R. Beene et al., *Possibility of pions emitted in spontaneous fission of* $^{252}Cf$, Phys. Rev. C 38 (1988) 569; [37] C. Cerruti, J. M. Hisleur, J. Julien, L. Legrain, Y. Cassagnou and M. Ribrag, *Search for pion emission from* $^{252}Cf$ *Source* Z.Phys. **A329**, 383 (1988).

[38] S. Stanislaus et al., *Search for spontaneous emission of pions*, Phys. Rev. C 39 (1989) 295;

[39] J. Jullien et al., *New Limit in Search for Pion Emission from* $^{252}Cf$ *Source*, Z. Phys. A332 (1989) 473;

[40] Yu. Adamchuk et al.*, Search for spontaneous emission of muons and/or charged pions*, Yad. Fiz. 49 (1989) 932.

[41] V. Bellini et al., Proceedings 14-th Europhys. Conf., *Rare Nuclear Processes*, Bratislava, 22-26 October 1990, (Ed.P.Povinec), p.116.

[42] K. Janko and P. Povinec, Proceedings 14-th Europhys. Conf., *Rare Nuclear Processes*, Bratislava, 22-26 October 1990, (Ed.Povinec), p.121.

[43] J. N. Knudson et al., *Search for neutral pions from the spontaneous fission of* $^{252}Cf$, Phys. Rev. C 44 (1991) 2869-2871

[44] H. Otsu et al., *Search for spontaneous* $\pi^-$ *emission from a* $^{252}Cf$ *source*, Z. Phys. A 342 (1992) 483;

[45] Khryachkov et al., *A Spectrometer for Investigation of Ternary Nuclear Fission*, Instr. Exp. Tech. **45,** 615 (2002).

[46] V.A.Varlachev et al., Russian Phys. Journ. **46,** 874 (2003).

[47] V.A.Varlachev et al., *Search for pi-zero emission from neutron-induced fission of 235-U nuclei,* Pisma JETP, **80,** 171 (2004).

[48] L. Arrabito et al., *Search for spontaneous muon emission from lead nuclei,* E-Preprint ArXiv hep-exp/0506078, June 2005, EXP. OPERA-Collaboration,CERN-Geneva.

[49] N.E. Holden and D.C. Hoffman, Pure Appl. Chem., **72,** 1525 (2000).

[50] J. F. Wild et al., Phys. Rev. **C32,** 488 (1985).

[51] C. A. Grady and R. M. Walker, Proceedings Intern. Symposium on Superheavy Elements, Lubbock, Texas, March 9-14, 1978, (Ed. M. A. K. Lodhi), Pergamon, New York (1978), p.191.

[52] G. G. Laemmlein, Nature, **155,** 724 (1945).